\documentclass{mem}
\usepackage{natbib}\usepackage{txfonts}\usepackage{balance}
\usepackage{graphicx}
\usepackage[a4paper]{hyperref}
\idline{80}{1}
\begin{document}
\newcommand{\ms}{\hbox{${\rm m\,s}^{-1}$}}
\newcommand{\kms}{\hbox{${\rm km\,s}^{-1}$}}
\newcommand{\zem}{\hbox{$z_{\rm em}$}}
\newcommand{\zabs}{\hbox{$z_{\rm abs}$}}
\newcommand{\SNR}{\hbox{${\rm SNR}$}}
\newcommand{\da}{\hbox{$\Delta\alpha/\alpha$}}
\newcommand{\bm}{\boldmath}

\def \lnp{Lecture~Notes~Phys.}
\def \sci{Science}

\defcitealias{GriestK_09a}{G09}
\defcitealias{MurphyM_03a}{M03}
\defcitealias{MurphyM_04a}{M04}

\title{
Keck constraints on a varying fine-structure constant: wavelength calibration errors
}

\subtitle{}

\titlerunning{Varying $\alpha$ from Keck}

\author{Michael T.\ Murphy\inst{1} \and John K.\ Webb\inst{2} \and Victor V.\ Flambaum\inst{2} }

\offprints{M.~T.~Murphy}
 
\institute{
Centre for Astrophysics and Supercomputing, Swinburne University of Technology, Melbourne, Victoria 3122, Australia; \email{mmurphy@swin.edu.au}
\and
School of Physics, University of New South Wales, Sydney NSW 2052, Australia
}

\authorrunning{Murphy}

\abstract{The Keck telescope's High Resolution Spectrograph (HIRES)
  has previously provided evidence for a smaller fine-structure
  constant, $\alpha$, compared to the current laboratory value, in a
  sample of 143 quasar absorption systems:
  $\da=(-0.57\pm0.11)\times10^{-5}$. The analysis was based on a
  variety of metal-ion transitions which, if $\alpha$ varies,
  experience different relative velocity shifts. This result is yet to
  be robustly contradicted, or confirmed, by measurements on other
  telescopes and spectrographs; it remains crucial to do so. It is
  also important to consider new possible instrumental systematic
  effects which may explain the Keck/HIRES results.
  \citet[][arXiv:0904.4725v1]{GriestK_09a} recently identified
  distortions in the echelle order wavelength scales of HIRES with
  typical amplitudes $\pm$250\,\ms. Here we investigate the effect
  such distortions may have had on the Keck/HIRES varying $\alpha$
  results. Using a simple model of these intra-order distortions, we
  demonstrate that they cause a random effect on \da\ from absorber to
  absorber because the systems are at different redshifts, placing the
  relevant absorption lines at different positions in different
  echelle orders. The typical magnitude of the effect on \da\ is
  $\sim$$0.4\times10^{-5}$ for individual absorbers which, compared to
  the median error on \da\ in the sample, $\sim$$1.9\times10^{-5}$, is
  relatively small.  Consequently, the weighted mean value changes by
  less than $0.05\times10^{-5}$ if the corrections we calculate are
  applied.  Unsurprisingly, with corrections this small, we do not
  find direct evidence that applying them is actually warranted.
  Nevertheless, we urge caution, particularly for analyses aiming to
  achieve high precision \da\ measurements on individual systems or
  small samples, that a much more detailed understanding of such
  intra-order distortions and their dependence on observational
  parameters is important if they are to be avoided or modelled
  reliably.   }
\maketitle{}

\section{Introduction}\label{sec:intro}

The Standard Model of particle physics is parametrized by several
dimensionless `fundamental constants', such as coupling constants and
mass ratios, whose values are not predicted by the Model itself.
Instead their values and, indeed, their constancy must be established
experimentally. If found to vary in time or space, understanding their
dynamics may require a more fundamental theory, perhaps one unifying
the four known physical interactions. One such parameter whose
constancy can be tested to high precision is the fine-structure
constant, $\alpha\equiv e^2/\hbar c$, characterising the strength of
electromagnetism. Earth-bound laboratory experiments, conducted over
several-year time-scales, which use ultra-stable lasers to compare
different atomic clocks based on different atoms/ions \citep[e.g.~Cs,
Hg$^+$, Al$^+$, Yb$^+$, Sr, Dy; e.g.][]{PrestageJ_95a}, have limited
the time-derivative of $\alpha$ to
$\dot{\alpha}/\alpha=(-1.6\pm2.3)\times10^{-17}{\rm \,yr}^{-1}$
\citep{RosenbandT_08a}

\begin{figure*}[t!]
\begin{center}
\includegraphics[width=0.70\textwidth]{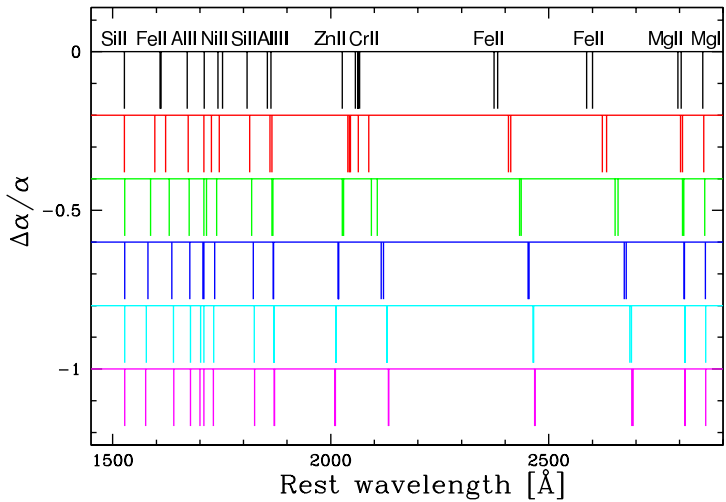}
\caption{\footnotesize Transitions used in MM analyses versus their
  rest-frame wavelength and their sensitivity to variations in
  $\alpha$. Note the very different signature of a varying $\alpha$ in
  absorption systems where only the Mg and Fe{\sc \,ii} transitions
  redwards of 2300\,\AA\ are detected and fitted, compared with the
  more complicated signature for systems containing a subset of the
  bluer transitions. Note that the relative strengths of the different
  transitions typically observed in quasar absorption systems is not
  portrayed here.}
\label{fig:shifts}
\end{center}
\end{figure*}

Important probes of variations over cosmological space- and
time-scales are narrow absorption lines imprinted on the spectra of
distant, background quasars by gas clouds associated with foreground
galaxies \citep{BahcallJ_67b}. In particular, electronic resonance
transitions from the ground states of metallic atoms and ions are
useful indicators of $\alpha$ variation for redshifts up to $\sim$4 --
when the universe was $\sim$10\,\% of its current age -- where the
transitions are easily accessed from ground-based optical telescopes.
The many-multiplet (MM) method \citep{DzubaV_99a,WebbJ_99a} utilises
the relative wavelength shifts expected from different transitions
from different multiplets of various atom/ions to measure $\alpha$
from quasar absorption spectra.  The velocity shift, $\Delta v_i$, of
transition $i$ due to a small relative variation in $\alpha$,
$\Delta\alpha/\alpha\ll 1$, is determined by the $q$-coefficient for
that transition,
\begin{equation}\label{eq:da1}
  \omega_{z,i} \equiv \omega_{0,i} + q_i\left[\left(\alpha_z/\alpha_0\right)^2-1\right]\,~{\rm or}
\end{equation}
\begin{equation}\label{eq:da2}
\frac{\Delta v_i}{c} \approx -2\frac{\Delta\alpha}{\alpha}\frac{q_i}{\omega_{0,i}}\,,
\end{equation}
where $\omega_{0,i}$ \& $\omega_{z,i}$ are the rest-frequencies in the
lab and at redshift $z$, $\alpha_0$ is the lab value of $\alpha$ and
$\alpha_z$ is the shifted value measured from an absorber at $z$. The
MM method is the comparison of measured velocity shifts from several
transitions (with different $q$-coefficients) to compute the best-fit
$\Delta\alpha/\alpha$. Figure \ref{fig:shifts} illustrates the
wavelength shifts experienced by the transitions typically utilized in
MM analyses.

Some evidence for $\alpha$-variation has emerged over the last decade
from quasar spectra observed with HIRES \citep{VogtS_94a} on the Keck
I 10-m telescope in Hawaii. The first tentative evidence in
\citet{WebbJ_99a} was subsequently strengthened with larger samples of
absorbers (\citealt{MurphyM_01a}; \citealt[hereafter
\citetalias{MurphyM_03a}]{MurphyM_03a}). MM analysis of 143 absorption
spectra, all from the Keck/HIRES instrument, currently indicate a
smaller $\alpha$ in the clouds at the fractional level
$\da=(-0.57\pm0.11)\times10^{-5}$ over the redshift range
$0.2<\zabs<4.2$ \citep[][hereafter
\citetalias{MurphyM_04a}]{MurphyM_04a}.

Obviously, confirmation of such a potentially fundamental result must
be made with many other telescopes and spectrographs. Similar MM
studies using the Ultraviolet and Visual Echelle Spectrograph on the
ESO Very Large Telescope in Chile are also beginning to yield
constraints on \da, but none yet rule out (or confirm) the Keck/HIRES
results \citep{MurphyM_08a}. At the same time, further searches for
subtle systematic errors which, despite extensive searches
\citep[e.g.][]{MurphyM_01b}, have evaded detection so far, must be
considered in detail. Of particular importance are systematic errors
in the wavelength calibration of the quasar spectra, which is usually
established using exposures of a standard thorium-argon (ThAr)
hollow-cathode emission-line lamp taken immediately before and/or after
the quasar exposure. After the wavelength--pixel mapping is derived
from the ThAr exposure, that solution is simply applied to the quasar
exposure. Since the quasar and ThAr light illuminate the spectrograph
slit differently and, in general, take slightly different paths
through the spectrograph to the CCD, wavelength calibration errors may
ensue.

\citet{GriestK_09a} (hereafter \citetalias{GriestK_09a}) recently
observed a quasar absorption system using Keck/HIRES with an iodine
gas absorption cell for more direct wavelength calibration, thereby
allowing comparison with the ThAr wavelength scale. For
varying-$\alpha$ studies, the most important and robust result from
\citetalias{GriestK_09a} is the identification of distortions in the
ThAr wavelength scale across each echelle order (hereafter
`intra-order distortions'). See their figures 4 and 5. These may
result from differential and variable vignetting of quasar light
compared to ThAr light, where only the former enters HIRES from the
telescope, the optical axes of which differ slightly, causing the
quasar light path to rotate as the telescope tracks the quasar
(\citealt{SuzukiN_03a}; \citetalias{GriestK_09a}).
\citetalias{GriestK_09a} also discuss velocity offsets between the
ThAr and I$_2$ wavelength scales. If those results are robust, they
are less important for varying $\alpha$ studies because, as equation
(\ref{eq:da2}) makes clear, such velocity offsets will not directly
influence a \da\ measurement\footnote{This is not strictly true when
  many quasar exposures are combined, as is generally the case for
  most of the Keck/HIRES sample. If different velocity shifts apply to
  the different exposures of the same quasar, and if the relative
  weights of the exposures vary with wavelength when forming the
  final, combined spectrum, then small relative velocity shifts will
  be measured between transitions at different wavelengths.
  Nevertheless, the effect on \da\ of overall velocity shifts is of
  secondary importance to the intra-order distortions considered in
  detail here.}.

This paper aims to assess the impact these intra-order distortions may
have had on the Keck/HIRES results for varying $\alpha$. The following
section describes the general effect the distortions will have and
provides a crude calculation of their expected magnitude. Section
\ref{sec:calc} details a more refined calculation of the correction to
\da\ for each absorber in the Keck/HIRES sample and Section
\ref{sec:res} discusses the results. In section \ref{sec:model} we
search for direct evidence for the need to apply the corrections and
consider the effect of model errors on our calculations. We conclude
in Section \ref{sec:conc}.

\section{The effect of intra-order wavelength distortions}\label{sec:distort}

The reason intra-order distortions are problematic for varying
$\alpha$ analyses is that they will produce velocity shifts {\it
  between} transitions. For an individual absorption system the
transitions will, in general, fall at different positions along
different echelle orders and will experience different velocity shifts
due to the wavelength calibration distortions. The
\citetalias{GriestK_09a} distortions generally cause blueward shifts
which increase towards the echelle order edges in either direction
from the center.  This pattern seems to be generally repeated on each
echelle order.  However, because the I$_2$ cell imprints absorption
lines on the quasar spectrum only over the wavelength range
$\sim$5000--6200\,\AA, no information about echelle orders outside
this range can be obtained. Nevertheless, it is clear that for two
absorbers at different redshifts, possibly with different transitions
detected and fitted, different spurious shifts in \da\ will be caused
by the intra-order distortions. Indeed, in general, the effect on \da\
will be random in sign and magnitude from absorber to absorber.

A simple illustrative estimate of the expected magnitude of the effect
on \da\ in a typical Mg/Fe{\sc \,ii} absorber can be calculated as
follows. Consider two transitions which, depending on the redshift of
the absorber, fall at different positions along their echelle order.
If we model the intra-order distortions found by
\citetalias{GriestK_09a} as a simple saw-tooth pattern of peak-to-peak
amplitude $\Delta v_{\rm amp}=500\,\ms$ then the root-mean-square
(RMS) velocity difference between the two transitions, averaged over
redshift, will be $\Delta v_{\rm amp}/(2\!\sqrt{2})=177\,\ms$ if the
transitions have arbitrary wavelengths. If one transition is from Mg
and the other is from Fe{\sc \,ii}, with typical rest wavelengths
$\sim$2700\,\AA\ (i.e.~frequencies $\sim$37000\,cm$^{-1}$), then the
typical difference in $q$-coefficients will be $\sim$1250\,cm$^{-1}$.
The resulting effect on \da\ then follows from equation
(\ref{eq:da2}): $\left|\da_{\rm corr}\right|\sim0.87\times10^{-5}$. Adding more
transitions will reduce this effect approximately as $\sqrt{N}$. Thus,
with typically 4--6 transitions observed in a single absorber, the
spurious effect the intra-order distortions will have on its value of
\da\ will be $\left|\da_{\rm corr}\right|\sim0.4\times10^{-5}$.

Degradation in the accuracy of the wavelength solution near the order
edges was already considered a possibility in \citet{WebbJ_99a}. To
test the possible effect of this \citeauthor{WebbJ_99a} artificially
increased the 1-$\sigma$ error bar on \da\ for absorption systems with
transitions falling near order edges, thereby down-weighting those
systems in the calculation of the weighted mean \da\ for the entire
sample. The effect was found to be insignificant. However, a more
detailed calculation is clearly warranted given the additional
information about the particular form of the intra-order wavelength
distortions identified by \citetalias{GriestK_09a}; we carry this out
in the following section.

\section{Calculating the effect on the Keck/HIRES results}\label{sec:calc}

In principle, the best way to understand the effect of intra-order
distortions on the Keck/HIRES results would be to correct the
wavelength scale of each extracted quasar exposure individually and
then combine together exposures of the same quasar to again form the
1-dimensional spectrum. The $\chi^2$ minimization of the Voigt profile
fits to the absorbers could be run again on the new versions of the
combined spectra, producing new, corrected values of \da. This,
clearly, would involve enormous effort and the results would still be
subject to possible errors in our model of the intra-order distortions
(hereafter `model errors'). We therefore take a cruder approach, but
one which still allows us to check our expectations that (a) the
correction to \da\ should be random in sign and magnitude from
absorber to absorber and (b) that the typical correction will be of
order $\da_{\rm corr}\approx0.4\times10^{-5}$ as calculated above
(Section \ref{sec:distort}).

We first assume an appropriate functional form to counter the
intra-order distortions of \citetalias{GriestK_09a}: a velocity shift
increasing from zero at the echelle order centre to $+$500\,\ms\
(i.e.~a redward shift) at the order edges linearly in wavelength
space. We consider a more complex model in Section \ref{sec:model} to
illustrate the likely size of model errors in our results.

For each absorption system in the Keck/HIRES sample, we must determine
where each transition falls with respect to the order edges to
estimate the velocity shifts between all the transitions due to the
intra-order distortions. This is complicated by the fact that most
quasar spectra used in the analysis were formed by combining extracted
spectra of many separate exposures. For each absorption system, we
derived the order positions for all relevant transitions from the
extracted, wavelength calibrated spectrum for each quasar exposure.
Thus, for each transition, $i$, an average velocity shift, $\Delta
v_i$, was determined assuming that all quasar exposures contributing
spectra of that transition did so with equal weight. While the
signal-to-noise ratio (\SNR) did vary between exposures, this is a
good approximation for most absorbers.

From the velocity shifts between the transitions in a given absorber,
equation (\ref{eq:da2}) can be used to compute the corresponding
correction to the value of \da. That is, $\da_{\rm corr}$ is the slope
(up to a factor of $-2$) of a linear least-squares fit of the values
of $\Delta v_i/c$ versus $q_i/\omega_{0,i}$. We use values for the
laboratory frequencies, $\omega_0$, and $q$-coefficients from
\citetalias{MurphyM_03a} in these calculations. However, two
complications arise here. The first is that not all transitions have
spectra of the same \SNR. This is easily remedied by weighting the
least squares fit with the squares of the \SNR\ values measured from
the combined spectra in the continuum around each transition.

The second complication is that the {\it shape} of the absorption
profile of each transition, and the relative {\it strength} of its
constituent velocity components, also determine its relative
contribution to constraining \da\ in a given absorber. For example, in
an absorber with many strong, marginally resolved velocity components,
a strong transition may appear very saturated. Thus, the centroids of
the velocity components are very weakly constrained, if at all. But if
several weaker transitions were also fitted, those velocity components
will be optically thin and their centroids will be strongly
constrained in the weak transitions. Thus, in this example, the strong
transition may not contribute much to the final constraint on \da\ in
this absorber. Properly taking this into account is as `simple' as
re-running the $\chi^2$ minimization of the Voigt profile fits to the
absorption profiles after correcting the spectrum of each transition
$i$ by its corresponding velocity shift, $\Delta v_i$. However, we
have ignored this complication, thereby allowing a simpler, more
illustrative calculation. Nevertheless, we have tested the importance
of this simplification by re-running the $\chi^2$ minimization on
several absorption systems, at both low- and high-$z$, and find it to
be fairly unimportant in most cases.

\section{Results}\label{sec:res}

\begin{figure*}[t!]
\begin{center}
\includegraphics[width=0.8\textwidth]{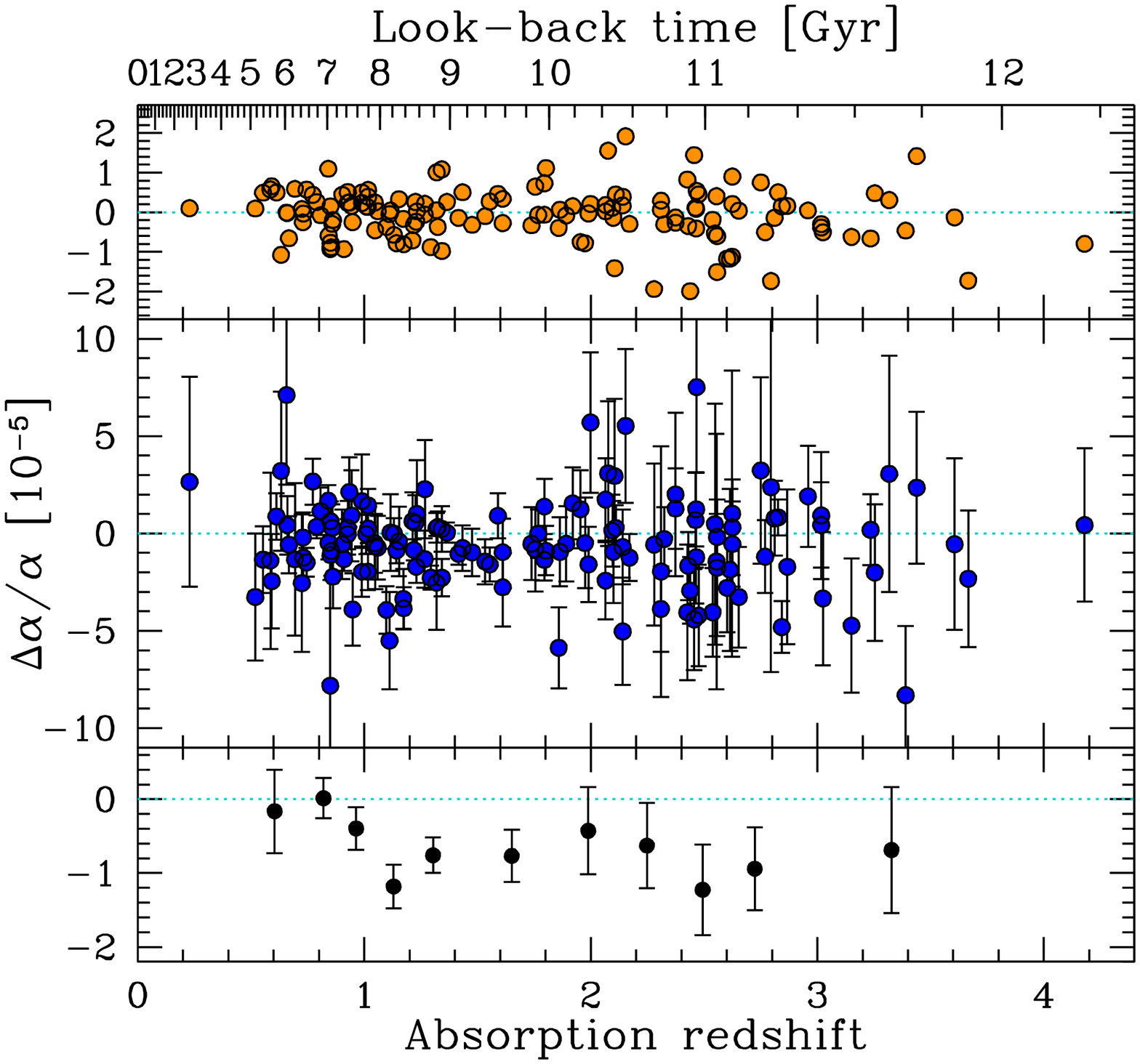}\vspace{-0.1em}
\caption{\footnotesize Corrections to \da\ and corrected values of
  \da\ versus absorption redshift and cosmological look-back time
  ($H_0=71\,\kms\,{\rm Mpc}^{-1}$, $\Omega_{{\rm m,}0}=0.27$,
  $\Omega_{\Lambda,0}=0.73$). {\it Top:} Estimated corrections to \da\
  due to intra-order distortions. Note that their sign and magnitude
  vary randomly from absorber to absorber. The median magnitude of the
  corrections is $0.37\times10^{-5}$. {\it Middle:} The corrected
  values of \da\ with their error bars representing the quadrature sum
  of the 1-$\sigma$ statistical and systematic errors. {\it Bottom}:
  Weighted mean values of \da, with 1-$\sigma$ errors, in bins
  containing 13 absorbers each.}
\label{fig:res}
\end{center}
\end{figure*}

Our estimates of the corrections to \da\ for each of the 143 absorbers
in the \citetalias{MurphyM_04a} Keck/HIRES sample, calculated using
the method described above, are given in Table \ref{tab:res}. These
corrections should be added to the original values of \da, which are
also given in the table for convenience. The upper panel of
Fig.~\ref{fig:res} shows the corrections plotted versus the redshifts
of the absorbers. Immediately we notice that the sign and magnitude of
the corrections vary randomly from absorber to absorber. Indeed, the
mean correction, $(-0.05\pm0.05)\times10^{-5}$, is consistent with
zero, indicating that there is no strong influence on the average
value of \da, as expected. The median magnitude of the corrections is
$0.37\times10^{-5}$, in line with expectations from the very crude
estimate made in Section \ref{sec:distort}. And while the assumptions
required to perform the calculation for each absorber, detailed in the
last section, are not unimportant, they should not significantly
affect these two main conclusions.

\begin{table*}
  \caption{\footnotesize Values of \da\ and corrections for possible intra-order distortions of the wavelength scale. The first three columns specify the B1950 quasar name, the emission and absorption redshifts. The fourth column provides the transitions used to determine \da\ in each absorber; each letter represents a transition and the key is provided in table 2 of \citetalias{MurphyM_03a}. The fifth column is the raw value of \da\ together with the formal 1-$\sigma$ statistical error. The systematic error term, $\delta(\da)_{\rm sys}$, is non-zero only for ``high-contrast'' absorbers; see text for explanation. The seventh and eighth columns provide our estimate of the correction, $\da_{\rm corr}$, for intra-order distortions, to be added to \da, and the systematic error term, $\delta(\da)^{\rm corr}_{\rm sys}$, which was calculated in the same way as $\delta(\da)_{\rm sys}$ but using the corrected values of \da. The final column specifies the observational sub-samples employed: A = ``Previous low-$z$ sample'' from \citetalias{MurphyM_03a}; B1 = ``Previous high-$z$ sample'' from \citetalias{MurphyM_03a}; B2 = The 15 absorbers added in \citetalias{MurphyM_04a}; C = ``New sample'' from \citetalias{MurphyM_03a}. Only a excerpt of the table is provided here; the complete table is available from the authors or http://astronomy.swin.edu.au/$\sim$mmurphy/pub.html\,.}\vspace{-2.5em}
\label{tab:res}
\footnotesize
\begin{center}
\begin{tabular}{lcccccccl}\hline
B1950 name   & \zem & \zabs   & Transitions  & \da                        & $\delta(\da)_{\rm sys}$ & $\da_{\rm corr}$ & $\delta(\da)^{\rm corr}_{\rm sys}$ & Sample \\
             &      &         &              & [$10^{-5}$]                & [$10^{-5}$]             & [$10^{-5}$]      & [$10^{-5}$]                        &        \\\hline
1634$+$7037  & 1.34 & 0.99010 & bcnpqr       & $\phantom{-}1.156\pm2.399$ & 0.000                   &  0.498           & 0.000                              & A      \\
0019$-$1522  & 4.53 & 3.4388  & ghl          & $\phantom{-}0.937\pm3.912$ & 0.000                   &  1.414           & 0.000                              & B1     \\
0100$+$1300  & 2.68 & 2.3095  & efgjklmvw    & $-3.949\pm1.370$           & 1.754                   &  0.063           & 1.707                              & B1     \\\hline
\end{tabular}
\end{center}
\normalsize
\end{table*}

It is important to emphasise that the sub-samples of absorbers below
and above $\zabs=1.8$ are qualitatively different. All the absorbers
at $\zabs<1.8$ (`low-$z$') depend only on Mg and Fe{\sc \,ii}
transitions, whereas those at $\zabs>1.8$ (`high-$z$') tend not to
include Mg and generally contain the wider variety of transitions,
with a mixture of $q$-coefficients, bluewards of $\lambda_{\rm
  rest}=2300$\,\AA\ in Fig.~\ref{fig:shifts}. As illustrated in
Fig.~\ref{fig:shifts}, the signature of a varying $\alpha$ for
Mg/Fe{\sc \,ii} absorbers is therefore very simple, while that for the
higher-$z$ systems is more complicated, strongly depending on which
transitions are detected and fitted. In this sense, absorbers with
only Mg/Fe{\sc \,ii} transitions fitted are more susceptible to simple
instrumental systematic errors which cause long-range, low-order
distortions of the wavelength scale and also astrophysical effects
which might shift velocity components of Fe{\sc \,ii} relative to
those of Mg. However, for the low-$z$ Mg/Fe{\sc \,ii} absorbers in the
Keck/HIRES sample, the scatter around the weighted mean \da\ was
consistent with the individual error bars in \citetalias{MurphyM_03a}
and \citetalias{MurphyM_04a}. This remains true even after the
corrections for intra-order distortions are applied: $\chi^2=76.03$
around the weighted mean \da\ for the 77 low-$z$ absorbers.

The same cannot be said of the high-$z$ absorbers: in
\citetalias{MurphyM_03a} and \citetalias{MurphyM_04a} we identified a
sub-sample of high-$z$ absorbers for which the scatter in \da\
exceeded expectations based on the individual errors. In these 27
absorbers both very strong and very weak transitions were fitted
simultaneously -- i.e.~they are ``high-contrast'' absorbers -- making
the multi-component Voigt profile fitting process more difficult and
error-prone. These systems could be `under-fitted' -- too few velocity
components used to model each absorber -- and this may lead to
systematic errors in individual absorbers which are random in sign and
magnitude. We demonstrated the effect of under-fitting with simulated
spectra in \citet{MurphyM_08a}. Just as in \citetalias{MurphyM_03a}
and \citetalias{MurphyM_04a}, the average systematic error in the
sample of 27 high-contrast absorbers is that which, when added in
quadrature to the individual statistical errors, reduces $\chi^2$ per
degree of freedom, $\chi^2_\nu$, to unity around the weighted mean
\da\ for those absorbers. The fact that we find a very similar
systematic error term using the corrected values of \da,
$1.71\times10^{-5}$, as found from the uncorrected values,
$1.75\times10^{-5}$, indicates that the extra scatter in \da\ for
these absorbers does not arise from the intra-order distortions; our
postulate that it arises from the mixture of strong and weak
transitions fitted, and the resulting under-fitting, remains. These
systematic error components are shown for the high-contrast absorbers
in Table \ref{tab:res}.

The middle panel of Fig.~\ref{fig:res} shows the corrected values of
\da\ with their 1-$\sigma$ statistical errors and systematic error
components added in quadrature. The lower panel shows a binned version
of these results, where the weighted mean \da\ of the 13 absorbers in
each bin is shown with its 1-$\sigma$ error. As with the uncorrected
results in \citetalias[see][]{MurphyM_04a}, \da\ is consistently
smaller $\alpha$ in the absorption systems compared to the current
laboratory value. Table \ref{tab:stats} quantifies this: the weighted
mean \da\ for the full sample, $\da=(-0.61\pm0.11)\times10^{-5}$,
differs only slightly from the uncorrected value from
\citetalias{MurphyM_04a}, $(-0.57\pm0.11)\times10^{-5}$, as expected.

\begin{table*}
  \caption{\footnotesize Basic statistics for the entire sample and redshift sub-samples. The second column gives the number of absorbers, $N_{\rm abs}$, in each sample. The statistics before any corrections for intra-order distortions are applied are in columns 3, 4 \& 5, while the remaining columns provide the same statistics after the \da\ values are corrected. $\left<\da\right>_{\rm w}$ is the weighted mean and $\left<\da\right>$ is the unweighted mean.}\vspace{-2.5em}
\label{tab:stats}
\footnotesize
\begin{center}
\begin{tabular}{lcccccccc}\hline
               &               & \multicolumn{3}{c}{Before correction}                            & & \multicolumn{3}{c}{After correction}                            \\ 
Sample         & $N_{\rm abs}$ & $\left<\da\right>_{\rm w}$ & $\left<\da\right>$  & Median        & & $\left<\da\right>_{\rm w}$ & $\left<\da\right>$  & Median       \\
               &               & [$10^{-5}$]                & [$10^{-5}$]         & [$10^{-5}$]   & & [$10^{-5}$]                & [$10^{-5}$]         & [$10^{-5}$]  \\\hline
{\bf Fiducial} & {\bf 143}     & \bm{$-0.57\pm0.11$}        & \bm{$-0.53\pm0.19$} & \bm{$-0.42$}  & & \bm{$-0.61\pm0.11$}        & \bm{$-0.58\pm0.20$} & \bm{$-0.56$} \\
$\zabs<1.8$    &  77           & $-0.54\pm0.12$             & $-0.61\pm0.22$      & $-0.39$       & & $-0.58\pm0.12$             & $-0.60\pm0.23$      & $-0.58$      \\
$\zabs>1.8$    &  66           & $-0.74\pm0.27$             & $-0.45\pm0.34$      & $-0.43$       & & $-0.79\pm0.27$             & $-0.57\pm0.34$      & $-0.55$      \\\hline
\end{tabular}
\end{center}
\normalsize
\end{table*}

Table \ref{tab:stats} gives the statistics for the low- and high-$z$
sub-samples. It was shown in \citetalias{MurphyM_03a} that the low-
and high-$z$ samples respond, on average, in opposite ways to simple,
long-range distortions of the wavelength scale. Note that the weighted
means for the low- and high-$z$ absorbers are similar and both depart
significantly from zero, even after the corrections are applied. This
also quantifies the fact that the evidence for a varying $\alpha$ is
dominated by the low-$z$ absorbers, with the evidence at high-$z$
being at the $\approx$3-$\sigma$ level both before and after
correction. Still, as mentioned above, if long-range wavelength
calibration distortions remain in the data, the low- and high-$z$
samples' weighted mean \da\ values should have opposite sign.  This is
an important internal consistency check that is only available when
one compares low-$z$ Mg/Fe{\sc \,ii} with higher-$z$ absorbers
constraining a greater diversity of transitions. Of course, we must
also recognise that absence of evidence for systematic errors capable
of explaining the consistency of the low- and high-$z$ \da\ values is
not evidence for their absence.

\section{Are the corrections warranted?}\label{sec:model}

We saw in the previous section that correcting the \da\ values for
intra-order distortions of the kind found by \citetalias{GriestK_09a}
makes no difference to the overall conclusions, nor for some more
detailed aspects of the Keck/HIRES results. So, having calculated
reasonable estimates of the corrections, can we find evidence that
applying them really is warranted?

For example, if the corrections are important, we should expect the
distribution of \da\ values around the mean to significantly narrow
after applying the corrections. We should also expect a significant
anti-correlation between the values of \da\ and the corrections. We
find neither of these effects.

Table \ref{tab:stats} provides the mean \da\ and its standard error --
i.e.~the RMS/$\sqrt{N_{\rm abs}}$ -- for the full sample and low- and
high-$z$ sub-samples. Only very small changes in the RMS are evident
in each case, indicating that the corrections do not remove a
significant amount of scatter in the \da\ values. A small Monte Carlo
simulation can be used to gauge how much difference in the RMS we
should expect if the corrections were important. We generated random
absorber data-sets, with the same size, total errors and corrections
as the real Keck/HIRES sample. The corrected \da\ values in each
realisation were set to $-0.6\times10^{-5}$ and then randomised,
according to the Gaussian errors for individual absorbers. The RMS of
this sample, and the same realisation with the corrections removed
from the \da\ values, were compared. Differences in RMS values between
the corrected and uncorrected realizations of $>0.1$ and
$>0.2\times10^{-5}$ occurred 37 and 4\,\% of the time by chance alone.
Thus, the lack of RMS differences between our real corrected and
uncorrected samples cannot be used as evidence that the corrections
are not meaningful. But the Monte Carlo simulation does indicate that
the corrections are too small, in comparison to the total errors on
\da\ in individual absorbers, to make a large difference to the sample
overall.

Similarly, the Spearman rank correlation and Kendall's $\tau$ tests
find insignificant anti-correlations between \da\ and the corrections
for the full sample or sub-samples. Nor are absorbers with large \da\
errors masking an underlying anti-correlation; using only absorbers
with total errors (quadrature sum of statistical and systematic
errors) less than $3\times10^{-5}$, gives similarly insignificant
results.  However, again a small Monte Carlo simulation confirms that,
effectively, this is not unexpected given the small size of the
corrections relative to the larger total errors on individual \da\
values.

To summarise, while we do not find evidence that the corrections we
calculate need to be applied, it is their relative smallness in
general which precludes a clear test for this.  In essence, the formal
errors in the Keck/HIRES sample still dominate over the potential
systematic errors caused by the intra-order distortions identified by
\citetalias{GriestK_09a}.

However, an important assumption so far has been the particular form
of intra-order distortion we have used; the \citetalias{GriestK_09a}
results have been approximated with a simple saw-tooth pattern in
every echelle order. Because the \citetalias{GriestK_09a} analysis
relies on I$_2$-cell calibration, it can only probe the intra-order
distortions over the wavelength range $\sim$5000--6200\,\AA. Even over
that relatively short wavelength range, there is some evidence for a
slow decrease in the peak-to-peak amplitude of the saw-tooth pattern
in bluer orders. How this extrapolates to longer and shorter
wavelengths than 6200 and 5000\,\AA\ (respectively) is unknown. Thus,
model errors may well exist in our estimates of the corrections to the
Keck/HIRES \da\ values above.

To test the importance of this, we can repeat the calculation of the
corrections using a somewhat different model of the intra-order
distortions: we fix $\Delta v_{\rm amp}$ to 500\,\ms at 5500\,\AA\ and
increase (decrease) it linearly with slope 0.55\,\ms\,\AA$^{-1}$ above
(below) 5500\,\AA\ and enforce a maximum (minimum) amplitude of
1000\,\ms\ (200\,\ms) at redder (bluer) wavelengths outside the range
covered by the I$_2$ cell calibration. The RMS difference between the
old and new corrections is $0.56\times10^{-5}$ and, with the new
corrections, the weighted mean becomes $(-0.62\pm0.11)\times10^{-5}$
over the whole sample and $(-0.61\pm0.12)\times10^{-5}$
[$(-0.66\pm0.27)\times10^{-5}$] at low$-z$ [high-$z$]. Comparison with
the values in Table \ref{tab:stats} reveals that the model errors in
this case are very small. Of course, this model is not very different
to our original, simpler one. For example, it may be that the
intra-order distortions have a completely different shape, phase
and/or amplitude for different quasar observations. This possibility
must be explored with future observations.

\section{Conclusions}\label{sec:conc}

Exploring systematic effects which may explain the Keck/HIRES evidence
for a varying $\alpha$ is clearly an important problem. However,
various properties of the Keck/HIRES results make the task difficult;
the consistency between the average \da\ values in the low-$z$
Mg/Fe{\sc \,ii} and the more diverse high-$z$ systems being an
important one. We have demonstrated here that if we model the
intra-order distortions identified by \citetalias{GriestK_09a} in a
simple way, and extrapolate the model to all echelle orders (not just
those within the I$_2$-cell calibration range) the effect on the
overall evidence for varying $\alpha$ is very small. As expected, the
distortions affect individual values of \da\ randomly in sign and
magnitude, because the differing redshifts place the transitions at
varying positions with respect to echelle order edges where the
distortions are worst. Indeed, the effect for a typical absorber is to
shift \da\ by $0.4\times10^{-5}$ which, compared to the median error
on \da, $\sim$$1.9\times10^{-5}$, is too small for us even to find
direct evidence of the need to apply the corrections we calculate.

Despite the above conclusions, it is important to emphasise that we do
not yet fully understand the origin of the intra-order distortions
identified by \citetalias{GriestK_09a}, how they depend on various
observational parameters (e.g.~telescope pointing direction,
temperature, time etc.) and therefore how they may differently affect
spectra of different quasars in the Keck/HIRES sample. Indeed,
\citetalias{GriestK_09a} show that separate exposures taken through
the I$_2$ cell seem to have somewhat different intra-order distortion
patterns. And while we have made a simple attempt to address this
problem of model errors, it is not enough to completely rule out
intra-order distortions as an important systematic error for the
Keck/HIRES results. If future Keck/HIRES observations allow much
smaller statistical errors on \da\ in individual absorbers or small
samples, intra-order distortions like those identified by
\citetalias{GriestK_09a} must either be eliminated or carefully
modelled.
 
\begin{acknowledgements}
  We thank K.~Griest and J.~B.~Whitmore for discussions. MTM thanks
  the Australian Research Council for a QEII Research Fellowship
  (DP0877998).
\end{acknowledgements}

\bibliographystyle{aa}

\end{document}